\documentclass[journal,twoside,web]{ieeecolor}
\usepackage{generic}
\usepackage{cite}
\usepackage{amsmath,amssymb,amsfonts,dsfont,mathrsfs}
\usepackage{empheq}
\usepackage{algorithm}
\usepackage{algorithmic}
\usepackage{graphicx}
\usepackage{multirow}
\usepackage{cases}
\usepackage{float}
\usepackage{setspace}
\usepackage{epstopdf}
\usepackage{diagbox}
\usepackage{subcaption}
\usepackage{marginnote}

\usepackage{boxedminipage}
\usepackage{booktabs}
\newcommand{\bremark}{ \begin{remark} \begin{rm} }
\newcommand{\eremark}{ \end{rm} \hfill $\triangleleft$ \end{remark} }
\newcommand{\btheorem}{ \begin{theorem} \begin{rm} }
\newcommand{\etheorem}{ \end{rm} \hfill $\triangleleft$ \end{theorem} }
\newcommand{\blemma}{ \begin{lemma} \begin{rm} }
\newcommand{\elemma}{ \end{rm} \hfill $\triangleleft$ \end{lemma} }
\newcommand{\bcorollary}{ \begin{corollary} \begin{rm} }
\newcommand{\ecorollary}{ \end{rm}  \end{corollary} }
\newcommand{\bdefinition}{ \begin{definition}\begin{rm} }
\newcommand{\edefinition}{ \end{rm} \hfill $\triangleleft$ \end{definition} }
\newcommand{\bproposition}{ \begin{proposition} \begin{rm} }
\newcommand{\eproposition}{ \end{rm}  \end{proposition} }
\newcommand{\bexample}{ \begin{example} \begin{rm} }
\newcommand{\eexample}{ \end{rm} \hfill $\triangleleft$ \end{example} }
\newcommand{\bproblem}{ \bf Problem \begin{rm} }
\newcommand{\eproblem}{ \end{rm} \hfill $\triangleleft$ \end{problem} }
\newcommand{\bproof}{ \textit{Proof:} \begin{rm} }
\newcommand{\eproof}{ \end{rm} \hfill $\square$}
\newcommand{\black}[1]{{\color{black}#1}}

\newtheorem{theorem}{\it Theorem}
\newtheorem{lemma}{\it Lemma}
\newtheorem{definition}{\it Definition}
\newtheorem{remark}{\it Remark}
\newtheorem{corollary}{\it Corollary}
\newtheorem{proposition}{\it Proposition}
\newtheorem{example}{\it Example}
\newtheorem{assume}{\it Assumption}

\newtheorem{problem}{\it Problem}

\def\BibTeX{{\rm B\kern-.05em{\sc i\kern-.025em b}\kern-.08em
    T\kern-.1667em\lower.7ex\hbox{E}\kern-.125emX}}
\begin{document}
\title{Bootstrap Policy Iteration for Stochastic Linear Quadratic Tracking with Multiplicative Noise}
\author{Jiayu Chen, Zhenhui Xu, and Xinghu Wang
\thanks{Jiayu Chen and Zhenhui Xu are with the Department of Systems and Control Engineering, Institute of Science Tokyo, Tokyo, 152-8552, Japan. (e-mail: jiayuc@lim.sc.e.titech.ac.jp; xu.z.al@m.titech.ac.jp.)}
\thanks{Xinghu Wang is with the Department of Automation, University of Science and Technology of China, Hefei, 230027, China (e-mail: xinghuw@ustc.edu.cn).}}

 \maketitle
 
\begin{abstract}
This paper studies the linear quadratic tracking problem for continuous-time stochastic  systems with multiplicative noise. 
The proposed framework formulates the problem under an average cost criterion and separates the computation of the optimal feedback and feedforward gains. 
By developing a bootstrap policy iteration algorithm, we eliminate the restrictive \textit{a priori} requirement for an initial mean square stabilizing feedback gain in existing policy iteration methods.
Based on this iterative framework, an off-policy reinforcement learning algorithm is proposed to learn the optimal feedback gain directly from data.
Using the learned feedback gain, the feedforward gain is subsequently obtained through a data-driven one-shot computation procedure.
These components work together to provide a model-free solution to the stochastic optimal tracking control problem.
The effectiveness of the proposed method is demonstrated through a numerical example.  
\end{abstract}

\section{Introduction}

\label{sec:introduction}
Reinforcement learning (RL), particularly in the form of adaptive dynamic programming, has emerged as a powerful data-driven framework for solving optimal control problems \cite{bertsekas2019reinforcement}.
\black{Its appeal lies in the ability to learn optimal control policies from input--state trajectory data, which is particularly useful when accurate system models are unavailable or costly to identify.}
Two fundamental approaches in RL are policy iteration (PI) \cite{kleinman1968iterative,jiang2012computational} and value iteration (VI) \cite{liu2012neural,bian2016value}.
PI generally exhibits fast, often quadratic, convergence, but it requires an initial stabilizing control policy, which is difficult to obtain when the system dynamics are unknown. VI removes this requirement, but typically converges more slowly.
Motivated by this tradeoff, several improved iterative schemes, including $\lambda$-PI algorithm, homotopy-based PI algorithm, and hybrid or composite methods, have been developed \cite{jiang2022bias,fan2025homotopy,gao2022resilient,ma2025adaptive}.
\black{Within this line of research, the linear quadratic tracking problem (LQTP) is widely used as a benchmark for developing and evaluating RL-based control methods.
This is because the LQTP is theoretically tractable while capturing reference-tracking objectives that arises in many practical systems, such as space manipulator systems \cite{wu2023space}, assistive human–robot interaction systems \cite{modares2015optimized}, and UAV systems \cite{karapetyan2023online}.
Its objective is to design a controller that drives the system output or state to follow a prescribed reference trajectory while minimizing a quadratic cost on the tracking error and control effort.}
A classical RL-based strategy for solving this problem involves constructing an augmented system by combining the plant dynamics with the command generator, and formulating an equivalent linear quadratic regulation problem with a discounted cost functional \cite{modares2014linear}. This reformulation enables existing RL regulation techniques to be applied directly to the tracking setting \cite{modares2014optimal,modares2016optimal1}.

\black{Despite these developments, most existing methods are designed for deterministic systems.
In many engineering applications, however, the system dynamics are subject to uncertainties, the magnitudes of which often vary according to the system state or control input, giving rise to stochastic systems with multiplicative noise \cite{zhao2019modeling,yang2013optimal}. Extending
RL-based control methods to such systems remains relatively underdeveloped, mainly because multiplicative noise complicates both stability analysis and optimality characterization. To address these challenges, several recent studies have developed RL algorithms for stochastic linear quadratic control problems \cite{li2022stochastic,li2024policy,xu2025mean}.
These methods incorporate moment-related quantities into the learning equations in place of  exact dynamical model information. In practical implementation, these quantities are estimated from sampled input–state trajectories.} More recently,
the stochastic linear quadratic tracking problems (SLQTPs) have also been investigated in \cite{zhang2025model} by extending the state-augmentation framework developed in \cite{modares2014linear}.

Nevertheless, two key limitations remain. 
\black{First, existing PI-based methods for stochastic systems require an initial mean square stabilizing feedback gain, which is particularly difficult to obtain when the system dynamics are unknown. To the best of our knowledge, how to eliminate this strong requirement for stochastic systems with multiplicative noise remains an open problem.
Second, the state augmentation technique for SLQTPs has limited generality across reference trajectories.
Since the desired system dynamics is embedded in the augmented system, the learned tracker is tied to a specific reference trajectory.
Therefore, changing the reference requires the entire tracker to be retrained.
Moreover, state augmentation substantially increases the dimension of the learning problem, especially for high-dimensional reference systems.}

Motivated by these observations, this paper investigates the continuous-time SLQTP with multiplicative noise. 
We first formulate the problem under an average cost criterion and then characterize the optimal tracker. 
In contrast to \cite{zhang2025model}, where the feedback gain (FBG) and feedforward gain (FFG) are computed simultaneously from an augmented algebraic Riccati equation, our formulation allows the two gains to be determined separately. In particular, the optimal FBG is computed by solving a stochastic algebraic Riccati equation (SARE) in the original system dimension, whereas the optimal FFG is obtained from a Sylvester equation.
This avoids the dimension increase caused by state augmentation and eliminates the need to relearn the FBG when the reference trajectory changes.
We then develop a new PI algorithm, termed bootstrap policy iteration (B-PI), for learning the FBG. Unlike conventional PI methods, which require an \textit{a priori} mean square stabilizing gain \cite{li2022stochastic,xu2025mean}, B-PI initializes the learning process with the trivial zero matrix gain. It first constructs a mean square stabilizing FBG within finite iterations and subsequently converges to the optimal solution. Based on this framework, we further develop an off-policy RL algorithm that learns the FBG without requiring exact knowledge of the system dynamics. Finally, once the FBG has been obtained, the corresponding FFG is computed directly using a data-based one-shot computation method.


The remainder of this paper is organized as follows.
Section \ref{sec2} formulates the SLQTP and derives its optimal solution.
Section \ref{sec3} presents the proposed B-PI framework and its convergence analysis, and develops the corresponding model-free tracking control method.
Section \ref{sec4} gives a numerical example, and Section \ref{sec5} concludes the paper.

The notation used throughout this paper is summarized as follows. Let $A =[a_{ij}]\in\mathbb{R}^{m\times n}$ and $R=[r_{ij}]\in\mathbb{S}^{m}$ be a symmetric matrix. We define the following operators: $\|A\|_R^2 := A^{\top}RA$; $\mathrm{vec}(A):=[a_{11},\cdots,a_{m1},a_{12},\cdots,a_{m2},$ $\cdots,a_{1(n-1)},\cdots,$ $a_{m(n-1)},a_{1n},\cdots,a_{mn}]\in\mathbb{R}^{nm}$; $\mathrm{vech}(R):=[r_{11},\cdots,r_{1m},r_{22},\cdots,r_{2m},\cdots,r_{(m-1)(m-1)},$  $r_{(m-1)m},r_{mm}]\in\mathbb{R}^{\frac{1}{2}m(m+1)}$.
Let $\{x(t)\in\mathbb{R}^n,t\geq0\}$ and $\{y(t)\in\mathbb{R}^m,t\geq0\}$ be  stochastic processes. For a given interval $T>0$, define $\delta_x(t):=\phi(x(t+T))-\phi(x(t))$ and $I_x(t):=\int_t^{t+T}\phi(x(\tau))\mathrm{d}\tau$, where $\phi(x)=\operatorname{vech}\big(2xx^\top-\operatorname{diag}(x)^2\big)$. Similarly, define $\delta{xy}(t):=x(t+T)\otimes y(t+T)-x(t)\otimes y(t)$ and $I_{xy}(t) := \int_{t}^{t+T}x(\tau)\otimes y(\tau)\mathrm{d}\tau$.
For a full column rank matrix $A$, $A^{\dagger} := (A^{\top} A)^{-1} A^{\top}$ denotes its Moore--Penrose inverse.
 
\section{Problem formulation and preliminaries}\label{sec2}
\subsection{Problem formulation}
Consider a continuous-time stochastic linear system 
\begin{subequations}\label{sys1}
\begin{empheq}[left={\Sigma:\empheqlbrace}]{align}
&\mathrm{d}x = (Ax+Bu)\mathrm{d}t+(Cx+Du)\mathrm{d}w, \label{sys1-1}\\
&y = Hx,\quad x(0)=x_0\in\mathbb{R}^n, \label{sys1-2}
\end{empheq}
\end{subequations}
where $x\in\mathbb{R}^n$, $u\in\mathbb{R}^m$, and $y\in\mathbb{R}^q$ represent the system state, control input, and system output, respectively. $w(t)$ is
a standard one-dimensional Brownian motion. The constant matrices $A, B, C,D, H$ have compatible dimensions and are assumed to be unknown.

The reference trajectory $y_d(t)$ is generated by 
\begin{equation}\label{sys2}
\Sigma_d:\left\{\begin{aligned}
&\dot{x}_d =A_dx_d,\\
&y_d = H_dx_d, \quad x_d(0)=x_{d0}\in\mathbb{R}^{n_d},
\end{aligned}\right.
\end{equation}
where $A_d \in \mathbb{R}^{n_d \times n_d}$ and $H_d \in \mathbb{R}^{q \times n_d}$ are also unknown.

\begin{assume}\label{A1}
System \eqref{sys1-1} or $\mathcal{S}:=[A,B|C,D]$ is mean square stabilizable, and the pair $[A,H|C]$ is exactly detectable.
\end{assume}

\begin{assume}\label{A2}
All eigenvalues of $A_d$ have zero real  parts, and their algebraic and geometric multiplicities are equal.
\end{assume}
\begin{remark}
The mean square stabilizability and exact detectability conditions in Assumption \ref{A1} are standard for analyzing stochastic systems \cite{zhang2008generalized,xu2025mean}.
\black{Assumption \ref{A2} allows the reference generator to produce a broad class of useful bounded signals, such as set-points \cite{xue2017add}, sinusoidal signals \cite{long2017active}, and multi-frequency trajectories \cite{lafmejani2020kinematic}. Compared with the case of an asymptotically stable reference generator, formulating an infinite-horizon optimal tracking problem for such non-decaying references is more technically demanding. In particular, since the tracking error and control input cannot approach zero simultaneously, the conventional infinite-horizon integral cost may become unbounded, and a well-posed formulation requires additional treatment, such as the discounted cost used in \cite{modares2014linear} or the average cost criterion adopted here.}
\end{remark}

The tracking performance is evaluated by the following cost 
\begin{equation}\label{P1}
P_d(u)=\lim_{T\rightarrow\infty}\frac{1}{T}J_{T}(u),
\end{equation}
where $J_{T}=\frac{1}{2}\mathbb{E}\left[\int_0^{T}(\|{y}-y_d\|_Q^2+\|{u}\|_R^2)\mathrm{d}t\right]$
with $ Q = Q^{\top}>0$ and $R=R^{\top} > 0$.
\begin{definition}[Admissible control policy]\label{def1}
A control policy is said to be admissible, denoted by $u\in\mathcal{A}$, if it has the form \black{$u(t)=-Kx(t)-Fx_d(t):=\mu(x(t),x_d(t))$} and satisfies the following properties:
{\romannumeral1}). system (\ref{sys1-1}) with $\mu(x,{0})$ is asymptotically mean square stable; {\romannumeral2}). $\lim \limits_{T\rightarrow\infty}\frac{1}{T}\mathbb{E}\left[\int_{0}^{T}\|x(\tau)\|^2\mathrm{d}\tau\right]<\infty$.
\end{definition}
 
\begin{problem}\label{prob1}
    For the plant (\ref{sys1}) and the reference system (\ref{sys2}), it is aimed to find $u^*\!\in\!\mathcal{A}$ such that $P_d(u^*)=\inf_{u\in\mathcal{A}}P_d(u)$.
\end{problem}

\subsection{Optimality analysis}
We derive the optimal solution of Problem \ref{prob1} by applying the calculus of variations directly. First, given matrices from $\mathcal{S}$ and a feedback gain matrix $K\in\mathbb{R}^{m\times n}$, we introduce the generalized Lyapunov operator $\mathcal{L}_{[K;\mathcal{S}]}:\mathbb{S}^n\rightarrow\mathbb{S}^n$ defined as 
\begin{equation*}
\begin{aligned}
\mathcal{L}_{[K;\mathcal{S}]}(X) :=& (A - BK)^{ \top} X + X(A - BK) \\
&+ (C - DK)^{ \top} X (C - DK)
\end{aligned}
\end{equation*}
with its spectrum given by
\begin{equation*}
\sigma(\mathcal{L}_{[K;\mathcal{S}]}) := \{\lambda\in\mathbb{C}:\mathcal{L}_{[K;\mathcal{S}]}(X)=\lambda X,~X\in\mathbb{S}^n,~X\neq0\}.
\end{equation*}
Let $x^*(t)$ denote the optimal state trajectory associated with $u^*(t)$. We examine perturbed trajectories around this optimal pair $(u^*,x^*)$, taking the form ${u}(t)=u^*(t)+\epsilon\delta u(t)$ and $x(t)=x^*(t)+\epsilon\delta x(t)$, with small $\epsilon\in\mathbb{R}$. Here, $u(t)\in\mathcal{A}$ and $\delta x(t)$ evolves according to $\mathrm{d}\delta x=(A\delta x+B\delta u)\mathrm{d}t+(C\delta x+D\delta u)\mathrm{d}w$ with $\delta x(0)=0$. The cost functional \eqref{P1} expands as
\begin{equation}\label{eq:Taylor}
P_d(u)=P_d(u^*)+\lim_{T\rightarrow\infty}\frac{1}{T}\left( \epsilon\delta J_{T}+\frac{\epsilon^2}{2}\delta^2 J_{T}\right),
\end{equation}
where 
\begin{equation*}
   \begin{aligned}
        &\delta J_{T}=\mathbb{E}\left[\int_0^{T}  \left((Hx^* - H_dx_d)^{\top}QH \delta x+{u^*}^{\top}R\delta u \right) \mathrm{d}\tau \right],\\
        &\delta^2 J_{T}=\mathbb{E}\left[ \int_0^{T} \! \left(\delta x^{\top}H^{\top}QH \delta x+ \delta u^{\top}R\delta u\right) \mathrm{d}\tau\right].
    \end{aligned} 
\end{equation*}
Since $H^{\top} Q H \geq 0$ and $R>0$, it follows immediately that $\lim_{T\rightarrow\infty}\frac{1}{T}\frac{\epsilon^2}{2}\delta^2J_T\geq0$. 

To derive the first-order necessary conditions, introduce a costate process $\boldsymbol{p}:[0,\infty)\rightarrow\mathbb{R}^n$ satisfying
$d\boldsymbol{p}(t) = \boldsymbol{f}(t)\mathrm{d}t+\boldsymbol{q}(t)\mathrm{d}w(t)$, where $\boldsymbol{f}(t)$ and $\boldsymbol{q}(t)$ will be specified later. Consider  the term $\mathbb{E}[\int_{0}^{T}\mathrm{d}(\boldsymbol{p}(t)^{\top}\delta x(t))]$, which equals to $\mathbb{E}[\boldsymbol{p}(T)^{\top}\delta x(T)]$ due to $\delta x(0)=0$. Applying It\^{o}' s formula, we rewrite this term as $\mathbb{E}\left[\int_{0}^{T}\mathrm{d}(\boldsymbol{p}(t)^{\top}\delta x(t))\right]=\mathbb{E}[\int_{0}^{T}((\boldsymbol{f}+A^{\top}\boldsymbol{p}+C^{\top}\boldsymbol{q})^{\top}\delta x$ $ +(B^{\top}\boldsymbol{p}+D^{\top}\boldsymbol{q})^{\top}\delta u)\mathrm{d}t]$. Thus, we have the following equality $\mathbb{E}[\int_{0}^{T}((\boldsymbol{f} + A^{\top}\boldsymbol{p} + C^{\top}\boldsymbol{q})^{\top}\delta x +(B^{\top}\boldsymbol{p} + D^{\top}\boldsymbol{q})^{\top}\delta u)\mathrm{d}t]=\mathbb{E}\left[\boldsymbol{p}(T)^{\top}\delta x(T)\right]$ holds.  Substituting this into $\delta J_T$ yields	
\begin{equation}\label{dJ}
\begin{aligned}
\!\!\!\!\delta J_{T}\!=\!\mathbb{E}\left[\int_0^{T} ( f_1^{\top}\delta x\!+f_2^{\top}\delta u ) \mathrm{d}\tau\right]\!-\!\mathbb{E}\!\left[\boldsymbol{p}(T)^{\!\top}\!\delta x(T)\right],
\end{aligned}
\end{equation}	
where $f_1=(H^{\top}Q(Hx^*-H_dx_d)+\boldsymbol{f}+A^{\top}\boldsymbol{p}+C^{\top}\boldsymbol{q}\big)$ and $f_2=\left(Ru^*+B^{\top}\boldsymbol{p}+D^{\top}\boldsymbol{q}\right)$. Following the sweep method, the costate can be expressed as
\begin{equation}\label{eq:p}
\boldsymbol{p}(t)=P^*x^*(t)+\Pi^*x_d(t),
\end{equation}
leading to
\begin{align}
&\boldsymbol{f}(t) = P^*Ax^*(t)+P^*Bu^*(t)+\Pi^*A_dx_d(t),\label{eq:f}\\
&\boldsymbol{q}(t) = P^*(Cx^*(t)+Du^*(t)).\label{eq:q}
\end{align}
Additionally, the terminal term of \eqref{dJ} satisfies
\begin{equation}\label{eq:terminal condition}
    \begin{aligned}
        &\epsilon\left|\mathbb{E}\left[ \boldsymbol{p}(T)^{\top}\delta x(T)\right]\right|\leq \frac{\|P^*\|_2}{2}\left(3\|x^*(T)\|_2^2+\|x(T)\|_2^2\right)\\
        &~~~~~~~~~~~~~~~~~~~~+\frac{\|\Pi^*\|_2}{2}\left(\|2\|x_d(T)\|_2^2+x(T)\|_2^2\right)
    \end{aligned}
\end{equation}
Since $u(t),u^*(t)\in\mathcal{A}$, and $x_d(t)$ is bounded, this gives rise to $\lim_{T\rightarrow\infty}\frac{\epsilon}{T} \mathbb{E}\left[ \boldsymbol{p}(T)^{\top}\delta x(T)\right]=0$.

Setting the first variation to zero for all admissible perturbations implies $f_1\!=\!{0}$ and $f_2\!=\!{0}$. By substituting Eqs. \eqref{eq:p} and \eqref{eq:q} into $f_2=0$, we obtain $(R+D^{\top}P^*D)u^*(t) +(B^{\top}P^*+D^{\top}P^*C)x^*(t)+B^{\top}\Pi^*x_d(t)=0$. If the matrix $R+D^{\top}P^*D$ is inverse, then the optimal control policy is uniquely determined by
\begin{equation}\label{eq:u*}
u^*(t)=-K^*x^*(t)-F^*x_d(t),  
\end{equation}
where
\begin{align}
&K^*=(R+D^{\top}P^*D)^{-1}(B^{\top}P^*+D^{\top}P^*C),\label{eq:K*}\\
&F^*=(R+D^{\top}P^*D)^{-1}B^{\top}\Pi^*,\label{eq:F*}
\end{align}
Further substituting Eqs. \eqref{eq:p}-\eqref{eq:q} and the obtained optimal control policy into the condition $f_1=0$ leads to the equality $(\mathcal{L}_{[K^*;S]}(P^*)+\|K^*\|_R^2+\|H\|_Q^2)x^*+(\Pi^*A_d+(A-BK^*)^{\top}\Pi^*-H^{\top}QH_d)x_d=0$, which holds for all $[{x^*}^{\top},x_d^{\top}]^{\top}\in\mathbb{R}^{n+n_d}$. Consequently, we deduce the following euquations 
\begin{equation}\label{eq:P*}
\begin{aligned}
\!\!\!\!&A^{\top}P^*+P^*A+C^{\top}P^*C-(P^*B+C^{\top}P^*D)\\
\!\!\!\!&\times(R+D^{\top}P^*D)^{-1}(B^{\top}P^*+D^{\top}P^*C)+\|H\|_Q^2=0,
\end{aligned}
\end{equation}
\begin{equation}\label{eq:Pi*}
\!\!\!\!\Pi^*A_d+(A-BK^*)^{\top}\Pi^*-H^{\top}QH_d=0.~~~~~~~~~~~ 
\end{equation}
\begin{theorem}
Suppose Assumptions \ref{A1} and \ref{A2} hold. Then \eqref{eq:P*} admits a unique stabilizing solution $P^*\geq0$, and \eqref{eq:Pi*} admits a unique solution $\Pi^*$. Moreover, the control policy given by \eqref{eq:u*} is an optimal solution for Problem \ref{prob1}, that is $P_d(u)\geq P_d(u^*),\forall u\in\mathcal{A}$.  

\end{theorem}
\bproof
Under Assumption \ref{A1}, by \cite[Theorem 4.1]{zhang2008generalized}, there exists a unique solution $P^*\geq0$ to Eq \eqref{eq:P*} such that $\sigma(\mathcal{L}_{[K^*;\mathcal{S}]})\subset\mathbb{C}^-$. It follows that the matrix $A_c:= A-BK^*$ is Hurwitz and the matrix $R+D^{\top}P^*D$ is positive definite, which ensure the uniqueness of solution to \eqref{eq:u*}. Moreover, Assumption 2 ensures that all the eigenvalues of $A_d$ have zero
real parts. Therefore, for all $i,j$, $\mathrm{Re}(\lambda_j(A_c))+\mathrm{Re}(\lambda_i(A_d))<0$, which implies that Eq. \eqref{eq:Pi*} admits a unique solution. We then conclude from \eqref{eq:u*} that $u^*\in\mathcal{A}$. 
Combing this with \eqref{dJ} and \eqref{eq:terminal condition}, we obtain $\frac{\epsilon}{T}\delta J_T(x^*)=0$
Finally, substituting this result into \eqref{eq:Taylor} yields $P_d(u)\geq P_d(u^*)$.  
\eproof
\begin{remark}
Note that the method in \cite{zhang2025model} yield an optimal tracker with a structure similar to \eqref{eq:u*}. However, in \cite{zhang2025model}, the FBG and FFG are computed jointly from an $(n+n_d)\times(n+n_d)$-dimensional SARE, whereas the proposed framework decouples their computation. Therefore, the FBG can be learned in the original $n$-dimensional state space,  
without introducing redundant learning parameters. 
Moreover, although $H_d$ is fixed for each instance of Problem \ref{prob1}, one may consider a family of tracking problems with identical plant dynamics, reference dynamics $A_d$, and cost functional, but different output matrices $H_d$. Since the SARE \eqref{eq:P*}, and hence $P^*$ and $K^*$, are independent of $H_d$, the learned FBG can be reused for all such problem instances, while the corresponding FFG is obtained for each $H_d$ through a one-shot computation.
\end{remark}

\section{Main results}\label{sec3}
In this section, we present a model-free approach to optimal tracking control. At its core is a novel, two-phase, iterative framework known as \emph{bootstrap policy iteration}, which is used to compute the optimal FBG. The term \emph{bootstrap} refers to the self-starting stabilization mechanism, which enables iteration to proceed without the need for an \textit{a priori}  stabilizing FBG.

\subsection{Useful operators and lemmas }
To facilitate the development of both model-based and model-free methods, we first establish several key results.

Let $\gamma,\alpha>0$ be two auxiliary constants and define the shifted stochastic system 
\begin{equation*}
    \mathcal{S}_{(\alpha,\gamma)}:= [A_{(\alpha,\gamma)},B|C,D],\quad A_{(\alpha,\gamma)}:= A-\frac{1}{2}(\gamma-\alpha)I_n.
\end{equation*}
For this system, the set of mean square stabilizing FBGs for $\mathcal{S}_{(\alpha,\gamma)}$ is defined as
\begin{equation*}
    \mathcal{Z}_{(\alpha,\gamma)}:=\{K\in\mathbb{R}^{m\times n}| \sigma(\mathcal{L}_{[{K};S_{(\alpha,\gamma)}]})\subset\mathbb{C}^-\}. 
\end{equation*}
Note that when $\alpha=\gamma$, the shifted system coincides with the original system $\mathcal{S}$, 
and hence $\mathcal{Z}_{(\gamma,\gamma)}$ is the set of mean square stabilizing FBGs for $\mathcal{S}$.

For later use, consider the Lyapunov equation
\begin{equation}\label{Lya-alpha}
    \mathcal{L}_{[K;\mathcal{S}(\alpha,\gamma)]}({P}) +\|{K}\|_R^2+\mathcal{Q}=0,
\end{equation}
together with update operators 
\begin{empheq}[left={\left\{\begin{aligned}},right={\end{aligned}\right.}]{align}
&\mathcal{K}(P) := (R+D^{\top}PD)^{-1}(B^{\top}{P}+D^{\top}{P}C)\label{update-K},\\
&\mathcal{R}(\alpha,P,K,\mathcal{Q}):= \alpha+\eta\frac{\lambda_{\min}\left(\|K\|_R^2+\mathcal{Q}\right)}{\lambda_{\max}\left({P}\right)}, \label{update-alpha}
\end{empheq}
where $\mathcal{Q}\in\mathbb{S}^n$ is an auxiliary weighting matrix to be specified later, $\eta\in(0,1)$, and $\lambda_{\min}(\cdot)$ and $\lambda_{\max}(\cdot)$ denote the minimum and maximum eigenvalues of a symmetric matrix, respectively.

Let  $\Theta\in\mathbb{S}^n$ be a prescribed positive definite matrix. 
\begin{lemma}\label{lem:LE-discount}
 Suppose Assumption \ref{A1} holds. Let $K\in\mathcal{Z}_{(\alpha,\gamma)}$ and $P^*\geq0$ be the stabilizing solution of \eqref{eq:P*}. Let $K'=\mathcal{K}(P)$ and $\alpha'=\mathcal{R}(\alpha,P,K',\Theta)$. Then the following statements hold.
\begin{enumerate}
\item If $\mathcal{Q}=\Theta$ and $\alpha\in\mathcal(0,\gamma)$, then \eqref{Lya-alpha} has a unique solution $0<{P}< c I$ for some finite constant $c>0$ such that $K'\in\mathcal{Z}_{(\alpha',\gamma)}$ and $\alpha'>\alpha$.
\item If $\mathcal{Q}=\|H\|_Q^2$ and $\alpha=\gamma$, then \eqref{Lya-alpha} has a unique solution ${P}\geq P^*$ such that $K'\in\mathcal{Z}_{(\gamma,\gamma)}$.
\end{enumerate}
\end{lemma}
See the proof in Appendix \ref{sec:proof_of_lemma_LE-discount}.  

\begin{lemma}\label{lem:alpha_init}
If there exist $\gamma>\alpha_0>0$ such that $\gamma>\sigma(\mathcal{L}_{[0;\mathcal{S}]})+\alpha_0$, then $0\in\mathcal{Z}_{(\alpha_0,\gamma)}$.    
\end{lemma}
\bproof
 Given $\gamma>\sigma(\mathcal{L}_{[0;\mathcal{S}]})+\alpha_0$, we can get $\sigma(\mathcal{L}_{[0;\mathcal{S}_{(\alpha_0,\gamma)}]})=\sigma(\mathcal{L}_{[0;\mathcal{S}]})-(\gamma-\alpha_0)\subset\mathbb{C}^-$.  Therefore, $0\in\mathcal{Z}_{(\alpha_0,\gamma)}$.
\eproof
\begin{remark}
Lemmas \ref{lem:LE-discount}--\ref{lem:alpha_init} provide the basis for the B-PI algorithm developed later. The auxiliary parameters $\alpha$ and $\gamma$ are introduced to construct a self-starting stabilization procedure. Specifically, $\gamma$ determines the shifted system family, while $\alpha$ is used to gradually adjust the shift. The initial value $\alpha_0$ and $\gamma$ are selected according to Lemma \ref{lem:alpha_init}, so that zero matrix gain is stabilizing for the shifted system $\mathcal{S}(\alpha_0,\gamma)$. 
This provides a valid initial condition for the subsequent iterative procedure without requiring a mean square stabilizing gain for the original system.
In addition, $\mathcal{Q}$ serves as an auxiliary design matrix in the Lyapunov equation. Its choice depends on the phase of the B-PI algorithm. Specifically, $\mathcal{Q}$ will be chosen as $\Theta$ during the search for a mean square stabilizing gain, and as $\|H\|_Q^2$ during the phase of improving the gain toward the optimal gain.
\end{remark}

Furthermore, the shifted system not only relaxes the requirement for an initial stabilizing FBG but also is useful for a model-free implementation that does not require explicit knowledge of the system dynamics. To this end, define
\begin{equation}\label{eq:dis1}
\chi(t):=e^{-\tilde{\alpha}t}x(t),\nu(t):=e^{-\tilde{\alpha}t}u(t),\zeta(t) := e^{-\tilde{\alpha}t}y(t),
\end{equation}
where $\tilde{\alpha}=\frac{1}{2}(\gamma-\alpha_0)$. It follows from \eqref{sys1} that these discounted variables satisfy the dynamics of the shifted system $\mathcal{S}_{(\alpha_0,\gamma)}$:
\begin{equation}\label{sys:xtilde}
 \left\{\begin{aligned} 
&\mathrm{d}\chi=\left(A_{(\alpha_0,\gamma)}\chi+B\nu\right)\mathrm{d}t+\left(C\chi+D\nu\right)\mathrm{d}w,\\
&\zeta = H \chi,~~\chi(0)=x_0.
\end{aligned}\right.
\end{equation}
The expected integral quantities are then defined along these trajectories as  $\bar{I}_{\chi}(t)$, $\bar{I}_{\chi \nu}(t)$, $\bar{I}_{\nu}(t)$, $\bar{I}_{K\chi}(t)$ and $\bar{\delta}_{\chi}(t)$ over each interval $[t,t+T]$, where $\bar{(\cdot)} := \mathbb{E}[(\cdot)]$. 
Based on these quantities, define
\begin{equation}\label{eq:psi}
\varphi_{[\alpha,K]}(t):=\left[\begin{array}{c}
\bar{\delta}_{\chi}(t)+(\alpha-\alpha_0)\bar{I}_{\chi}(t)\\
-2\big( (I_n \otimes K)\bar{I}_{\chi}(t) +\bar{I}_{\chi \nu}(t)\big)\\
-\bar{I}_{\nu}(t)+\bar{I}_{K\chi}(t)\end{array}\right].
\end{equation}

Let $T_s>0$ be the sampling interval, and $t_k= t_1+(k-1)T_s$ be the sampling instant, where $t_1$ is the initial time for data collection. 
The corresponding data matrices are defined as $\varPhi(\alpha,K):= [\varphi_{[\alpha,K]}(t_1),\cdots,\varphi_{[\alpha,K]}(t_l)]^{\top}$, $\mathcal{I}_{\chi}:=[\bar{I}_{\chi}(t_1),\cdots,\bar{I}_{\chi}(t_l)]$, $\mathcal{I}_{\chi \nu}:=[\bar{I}_{\chi \nu}(t_1),\cdots,\bar{I}_{\chi \nu}(t_l)]$, $\mathcal{I}_{\nu}:=[\bar{I}_{ \nu}(t_1),\cdots,\bar{I}_{\nu}(t_l)]$, $\Delta_{\chi}:=[\bar{\delta}_{\chi}(t_1),\cdots,\bar{\delta}_{\chi}(t_l)]$. To use the data-based matrices, we need to meet the following condition.
\begin{assume}\label{assume:rank1}
There exists an integer $l_1>0$ such that 
\begin{equation}\label{eq:rank1}
\begin{aligned}
\mathrm{rank}\left(\mathcal{I}\right) = 0.5n(n\!+\!1)\! + \!nm \!+\!0.5m(m\!+\!1),~\forall l\geq l_1,
\end{aligned}
\end{equation}
where $\mathcal{I}=[\mathcal{I}_{\chi}^{\top},\mathcal{I}_{\chi \nu}^{\top},\mathcal{I}_{\nu}^{\top}]^{\top}$.
\end{assume}
\begin{lemma}\label{lem:rank}
Suppose Assumption \ref{assume:rank1} holds.  If $K\in\mathcal{Z}_{(\alpha,\gamma)}$, then the matrix $\varPhi(\alpha,K)$ has full column rank.
\end{lemma}
See the proof in Appendix \ref{lempf:rank}.
\begin{remark}
Lemma \ref{lem:rank} lays the foundation for subsequent model-free design.
Assumption \ref{assume:rank1} imposes a finite-data excitation condition consistent with the initial excitation (IE) condition defined in \cite[Definition 2]{dhar2022initial}.
Specifically, the full row rank condition in \eqref{eq:rank1} is equivalent to $\mathcal{I}\mathcal{I}^{\top}>0$, which ensures that the data collected over the finite interval $[t_1,t_l+T]$ are sufficiently informative to uniquely determine the unknown parameters in the subsequent data-based equations.
In practice, this rank condition can be achieved by injecting designed probing signals into the input channel \cite{jiang2012computational,xu2025mean}.
\end{remark}

\subsection{ Bootstrap policy iteration }\label{sec:B-PI}

We are now in a position to present  the B-PI algorithm, summarized in Algorithm \ref{alg:Model_CPI}. The algorithm operates in two phases:
Phase \uppercase\expandafter{\romannumeral1} in lines 2-7, and Phase \uppercase\expandafter{\romannumeral2} in lines 8-12. 
The convergence of Algorithm \ref{alg:Model_CPI} and the asymptotic mean square stability of the resulting closed-loop systems are established in Theorem \ref{thm:DPI}, with detailed proof given in Appendix \ref{sec:proof_of_theorem_ref_thm_dpi}. Define the index $\mathbb{I}:= \min\{i\in\mathbb{N}|\alpha_i\geq \gamma\}$.
\begin{theorem}\label{thm:DPI}
Suppose that Assumption \ref{A1} holds and there exist constants $\gamma>\alpha_0>0$ such that $\gamma>\sigma(\mathcal{L}_{[0;\mathcal{S}]}) +\alpha_0$. Let $P^*\geq0$ be the stabilizing solution of \eqref{eq:P*} and $K^*=\mathcal{K}(P^*)$. Let the sequences $\{(P_i,K_i)\}_{i=1}^{\infty}$ and $\{\alpha_i\}_{i=1}^{\mathbb{I}}$ be generated by Algorithm \ref{alg:Model_CPI}. Then the following properties are satisfied.
\begin{itemize}
\item[1)] $\mathbb{I}<\infty$ and $K_{\mathbb{I}}\in\mathcal{Z}_{(\gamma,\gamma)}$.
\item[2)] $\lim_{i\rightarrow\infty}P_i=P^*$ and $\lim_{i\rightarrow\infty}K_i=K^*$.
\end{itemize}
\end{theorem}
\begin{algorithm}[!htb]
\caption{Bootstrap Policy Iteration}
\label{alg:Model_CPI}
\setlength{\abovedisplayskip}{2pt}
\setlength{\belowdisplayskip}{2pt}
\setlength{\abovedisplayshortskip}{2pt}
\setlength{\belowdisplayshortskip}{2pt}
\begin{algorithmic}[1]
\setlength{\itemsep}{1pt}
\setlength{\parsep}{0pt}
\setlength{\parskip}{0pt}
\STATE \textbf{Initialization:} $K_0=0$, $\alpha_0>0$, $\gamma>\max\{\alpha_0,\sigma(\mathcal{L}_{[0;\mathcal{S}]})+\alpha_0\}$, \black{$\eta\in(0,1)$}, $\varepsilon>0$, $i=0$.
\REPEAT
\STATE $i \leftarrow i+1$
\STATE Solve $P_i$ from
\begin{equation}\label{eq:int_P_i}
\mathcal{L}_{[K_{i-1};\mathcal{S}_{(\alpha_{i-1},\gamma)}]}(P_i)+\|K_{i-1}\|_R^2+\Theta=0.
\end{equation}
\STATE Update $K_i$ by
\begin{equation}\label{eq:int_K_i}
K_i = \mathcal{K}(P_i).
\end{equation}
\STATE Update $\alpha_i$ by
\begin{equation}\label{eq:int_alpha_i}
\alpha_i= \mathcal{R}(\alpha_{i-1},P_i,K_i,\Theta).
\end{equation}
\UNTIL{$\alpha_{i}\geq\gamma$}
\REPEAT
\STATE $i \leftarrow i+1$
\STATE Solve $P_i$ from 
\begin{equation}\label{eq:P_i}
\mathcal{L}_{[K_{i-1};\mathcal{S}]}(P_i)+\|K_{i-1}\|_R^2+\|H\|_Q^2=0.
\end{equation}
\STATE Update $K_{i}$ by  \eqref{eq:int_K_i}.
\UNTIL {$\|P_i-P_{i-1}\|_F\leq \varepsilon$}
\end{algorithmic}
\end{algorithm}
\begin{remark}
Classical PI methods ({\sl e.g.} \cite{li2022stochastic,li2024policy,xu2025mean}) for continuous-time stochastic linear systems require an initial mean square stabilizing FBG. In contrast, the proposed B-PI framework removes this restrictive requirement by introducing $\alpha_0$ and $\gamma$. 
The initial conditions on these parameters in Algorithm \ref{alg:Model_CPI} are mild and can be satisfied by choosing $\gamma$ sufficiently large and $\alpha_0$ sufficiently small. Moreover, our design employs a tunable weighting matrix that decouples the \emph{stabilize‑first} stage from the task cost functional, thereby relaxing the positive definiteness requirement on $\|H\|_Q^2$.
\end{remark}

\subsection{Model-free optimal tracking control}
Recall \eqref{sys:xtilde} and apply It\^{o}'s formula to $\chi^{\top}P_i\chi$ to obtain
\begin{equation*}
\begin{aligned}
\mathrm{d}\left(\chi^{\top}P_i\chi\right)
&=2\chi^{\top}P_i\left(A_{(\alpha_0,\gamma)}\chi+B\nu\right)\mathrm{d}t+\left(C\chi+D\nu\right)^{\top}\\
&\times P_i\left(C\chi+D\nu\right)\mathrm{d}t+2\chi^{\top}P_i\left(C\chi+D\nu\right)\mathrm{d}w.
\end{aligned}
\end{equation*}
Let $\Lambda_i:= D^{\top}P_iD$ and $M_i:= (R+\Lambda_i)K_i$. We then substitute \eqref{eq:int_K_i} and  $\mathscr{L}_{[K_{i-1};\mathcal{S}_{(\alpha,\gamma)}]}(P_i)$ into the above equation to obtain
\begin{equation}\label{eq:LE}
\mathscr{F}(P_i,M_i,\Lambda_i,\alpha)=\chi^{\top}\mathscr{L}_{[K_{i-1};\mathcal{S}_{(\alpha,\gamma)}]}(P_i)\chi.
\end{equation}
where  
\begin{equation*}
\begin{aligned}
&\mathscr{F}(P_i,M_i,\Lambda_i,\alpha) := \mathrm{d}\left(\chi^{\top}P_i\chi\right)+(\alpha-\alpha_0)\chi^{\top}P_i\chi\mathrm{d}t\\
&-2\chi^{\top}M_i^{\top}\left(K_{i-1}\chi+\nu\right)\mathrm{d}t-\nu^{\top}\Lambda_i\nu\mathrm{d}t+(K_{i-1}\chi)^{\top}\\
&\times\Lambda_i(K_{i-1}\chi)\mathrm{d}t+2\chi^{\top}P_i\left(C\chi+D\nu\right)\mathrm{d}w.
\end{aligned}
\end{equation*}
By replacing $\alpha$ with $\alpha_{i-1}$ and applying \eqref{eq:int_P_i} from Phase \uppercase\expandafter{\romannumeral1} of Algorithm \ref{alg:Model_CPI} to the right-hand side of \eqref{eq:LE}, we obtain
\begin{equation}\label{eq:LE1} 
\mathscr{F}( P_i, M_i, \Lambda_i, \alpha_{i-1} ) = -\chi(t)^{\top} \left(\|K_{i-1}\|_R^2+\Theta \right)\chi(t).
\end{equation}
Similarly, replacing $\alpha$ with $\gamma$ and applying \eqref{eq:P_i} from Phase \uppercase\expandafter{\romannumeral2} of Algorithm \ref{alg:Model_CPI}, the expression becomes
\begin{equation}\label{eq:LE2}
\!\!\mathscr{F}( P_i, M_i, \Lambda_i, \gamma) \!=\! -\chi(t)^{ \top} \|K_{i-1}\|_R^2\chi(t) - \zeta(t)^{ \top} Q\zeta(t).
\end{equation}
Next, we integrate both sides of \eqref{eq:LE1} and \eqref{eq:LE2} along (\ref{sys:xtilde}) over $[t,t+T]$, take expectations, and use Kronecker product representation, to obtain
\begin{align}
&\psi_{i-1}(t)^{\top}\theta_i\!=\!-\bar{I}_{\chi}(t)^{\top}\!\mathrm{vech}(\|K_{i-1}\|_R^2\!+\!\Theta),\label{eq:DB1-DPI}\\
&\phi_{i-1}(t)^{\top}\!\theta_i\!=\!-\bar{I}_{\chi}(t)^{\top}\!\mathrm{vech}(\|K_{i-1}\|_R^2)\!-\!\bar{I}_{\zeta}(t)^{\!\top}\!\mathrm{vech}(Q),\label{eq:DB1-PI}
\end{align}
where $\theta_i=[\mathrm{vech}(P_i)^{\top},\mathrm{vec}(M_i)^{\top},\mathrm{vech}(\Lambda_i)^{\top}]^{\top}$, $\psi_{i}(t) = \varphi_{[\alpha_{i},K_i]}(t)$, $\phi_i(t) = \varphi_{[\gamma,K_i]}(t)$, and $\bar{I}_{\zeta}(t) =\mathbb{E}[I_{\zeta}(t)]$.\\
Based on the sampling points, we construct the matrices $\Psi_{i} := [\psi_{i}(t_1),\cdots,\psi_{i}(t_l)]^{\top}$, $\Phi_{i-1} := [\phi_{i}(t_1),\cdots,\phi_{i}(t_l)]^{\top}$, and $\mathcal{I}_{\zeta}:=[\bar{I}_{\zeta}(t_1),\cdots,\bar{I}_{\zeta}(t_l)]$. This leads to the matrix-form representations of \eqref{eq:DB1-DPI} and \eqref{eq:DB1-PI} as
\begin{align}
&\Psi_{i-1} \theta_i=-\mathcal{I}_{\chi}^{\top}\mathrm{vech}(\|K_{i-1}\|_R^2+\Theta),\label{eq:Psi}\\
&\Phi_{i-1} \theta_i=-\mathcal{I}_{\chi}^{\top}\mathrm{vech}(\|K_{i-1}\|_R^2)-\mathcal{I}_{\zeta}^{\top}\mathrm{vech}(Q).\label{eq:Phi}
\end{align}
Note that  $\Psi_i\!=\!\varPhi(\alpha_i,K_i)$ and $\Phi_i\!=\!\varPhi(\gamma,K_i)$. Based on Lemma~\ref{lem:rank} and Theorem~\ref{thm:DPI}, We can now formalize the off-policy RL algorithm within the B-PI framework. Specifically, lines 4-5 of Algorithm \ref{alg:Model_CPI} are replaced by lines 5-6 of Algorithm \ref{alg:Data_OT}, and lines 10-11 of Algorithm \ref{alg:Model_CPI} are modified as lines 11-12 of Algorithm \ref{alg:Data_OT}.

\begin{theorem}\label{thm:CPI_DB}
Suppose the conditions of Theorem \ref{thm:DPI} and Assumption \ref{assume:rank1} hold. Let $\{(P_i,K_i,\alpha_i)\}_{i=1}^{\mathbb{I}}$ be generated by iteratively solving \eqref{eq:int_P_i_DB}-\eqref{eq:int_alpha_i_DB} until $\alpha_i\geq\gamma$, and let $\{(P_i,K_i)\}_{i=\mathbb{I}+1}^{\infty}$ be generated by iteratively solving \eqref{eq:P_i_DB} and \eqref{eq:int_K_i_DB}. Then the following properties hold.
\begin{itemize}
\item[1)] $\mathbb{I}<\infty$ and $K_{\mathbb{I}}\in\mathcal{Z}_{(\gamma,\gamma)}$.
\item[2)] $\lim_{i\rightarrow\infty}P_i=P^*$ and $\lim_{i\rightarrow\infty}K_i=K^*$. 
\end{itemize}
\end{theorem}
\bproof
From the above derivation and Lemma \ref{lem:rank}, it is easy to show that solving the  equations \eqref{eq:int_P_i_DB}-\eqref{eq:int_alpha_i_DB} is equivalent to solving   \eqref{eq:int_P_i}-\eqref{eq:int_alpha_i}. Therefore, statement 1 follows directly from Theorem \ref{thm:DPI}. After a mean square stabilizing gain is obtained in Phase \uppercase\expandafter{\romannumeral1}, solving \eqref{eq:P_i_DB} and \eqref{eq:int_K_i_DB} is equivalent to the Phase \uppercase\expandafter{\romannumeral2} iteration in Algorithm \ref{alg:Model_CPI}. Hence, the convergence result in statement 2 also follows from Theorem \ref{thm:DPI}.
\eproof

After obtaining the optimal $K^*$ and $\Lambda^*:=D^{\top}P^*D$, we proceed to develop a data-based equation for computing $F^*$.
Define the expected values $\bar{\chi} = \mathbb{E}(\chi), \bar{\nu} = \mathbb{E}(\nu), \bar{\zeta} = \mathbb{E}(\zeta)$. 

Using \eqref{eq:F*} and \eqref{eq:Pi*}, the time derivative of $\bar{\chi}(t)^{\top}\Pi^*x_d(t)$ can be expressed as
\begin{equation}\label{eq:0509-1}
\begin{aligned}
&\frac{\mathrm{d}}{\mathrm{d}t}(\bar{\chi}^{\top}\Pi^*x_d)=\bar{\chi}^{\top}H^{\top}QH_dx_d+(K^*\bar{\chi}+\bar{\nu})^{\top}\\
&~~~~~~\times(R+\Lambda^*)F^*x_d-\frac{1}{2}(\gamma-\alpha_0)\bar{\chi}^{\top}\Pi^*x_d.
\end{aligned}
\end{equation}
By integrating \eqref{eq:0509-1} over $[t,t+T]$ along the dynamics of $x_d$ and $\bar{\chi}$, and rearranging terms, we obtain 
\begin{equation}\label{eq:le-xi}
\begin{aligned}
\xi(t)^{\top} \vartheta^* = {I}_{y_d\bar{\zeta}}^{\top}\mathrm{vec}(Q), 
\end{aligned}
\end{equation}
where $\vartheta^*=[\mathrm{vec}(\Pi^*)^{\top},\mathrm{vec}(F^*)^{\top}]^{\top}$, and $\xi(t)$ is given by 
$$\xi = \left[\begin{array}{c}
\delta_{x_d\bar{\chi}}+\frac{1}{2}(\gamma-\alpha_0){I}_{x_d\bar{\chi}}\\
\!\!\!-(I_{n_d} \otimes(R + \Lambda^*){K^*}){I}_{x_d\bar{\chi}} - (I_{n_d} \otimes(R + \Lambda^*)){I}_{x_d\bar{\nu}}\!\!\!
\end{array}\right].$$
To facilitate a matrix-based formulation, we define the data-based matrices, ${\Delta}_{x_d\bar{\chi}} \!:=\![{\delta}_{x_d\bar{\chi}}(t_1),\!\cdots\!,\!{\delta}_{x_d\bar{\chi}}(t_l)]$, $\mathcal{I}_{x_d\bar{\chi}} \!:=\![I_{x_d\bar{\chi}}(t_1),\!\cdots\!,\!I_{x_d\bar{\chi}}(t_l)]$, $\mathcal{I}_{x_d\bar{\nu}}\!:=\![I_{x_d\bar{\nu}}(t_1),\!\cdots\!,\!I_{x_d\bar{\nu}}(t_l)]$, and $\mathcal{I}_{y_d\bar{\zeta}}\!:=\![I_{y_d\bar{\zeta}}(t_1),\!\cdots\!,\!I_{y_d\bar{\zeta}}(t_l)]$. We then construct the matrix $\Xi \!:=\![\xi(t_1),\cdots\!,\xi(t_l)]^{\top}$, which has the following structure 
\begin{equation*}
\Xi^{\top} \!=\!\left[\begin{array}{c}
\Delta_{x_d\bar{\chi}}+\frac{1}{2}(\gamma-\alpha_0)\mathcal{I}_{x_d\bar{\chi}}\\
\!\!\!(I_{n_d} \otimes (R+\Lambda^*){K^*})\mathcal{I}_{x_d\bar{\chi}} + (I_{n_d} \otimes (R+\Lambda^*))\mathcal{I}_{x_d\bar{\nu}} \!\!\! 
\end{array}\right].
\end{equation*}
Therefore, the linear matrix form of \eqref{eq:le-xi} becomes
\begin{equation}\label{eq:FF}
\Xi \vartheta^* = \mathcal{I}_{y_d\bar{\zeta}}^{\top}\mathrm{vec}(Q). 
\end{equation}
\begin{lemma}\label{lem:rank-xi}
If there exists $l_1>0$ such that for all $l\geq l_1$ the following rank condition holds
\begin{equation}\label{eq:rank2}
\mathrm{rank}([\mathcal{I}_{x_d\bar{\chi}},\mathcal{I}_{x_d\bar{\nu}}])=(n+m)n_d,
\end{equation}
then $\Xi$ has full column rank.
\end{lemma}

The proof is similar to that of Lemma \ref{lem:rank} and is omitted for brevity. As a result, under this condition, the unknown parameters can be uniquely determined by solving \eqref{eq:FF_DB}.
\begin{remark}
Algorithm \ref{alg:Data_OT} consists of two stages, {\sl i.e.,} FBG off-policy RL and FFG one-shot computation. This separation has two advantages. First, once the FBG is learned, different reference trajectories generated by the same exosystem can be tracked by recomputing only the corresponding FFGs, as illustrated in the numerical example. Second, compared with the augmentation formulation \cite{modares2014linear,zhang2025model}, the proposed approach reduces the storage burden by eliminating the redundant $(n_d\!\times\!n_d)$ block in the associated SARE. 
\end{remark}
\begin{remark}
Moreover, let $N_s$ denote the number of trajectory samples used for expectation estimation, $l$ the number of data intervals, and $h$ the number of discretization points used to approximate each integral. Because every integral is evaluated by a finite summation over $h$ points for each trajectory and each data interval, the dominant time complexity of Algorithm \ref{alg:Data_OT} is $O(h l N_s)$.
\end{remark}
{\color{black}\begin{remark}
In the practical implementation of Algorithm \ref{alg:Data_OT}, the expectations are approximated from sampled trajectories. Consequently, the variance of the observations appear through the finite-sample approximation errors in the data-based equations and affects
the accuracy of the learned parameters. Since these finite-sample errors propagate through the iterative updates, robustness against such perturbations is an important practical consideration. A related analysis can be found in \cite{xu2025robust}, showing that under bounded disturbances, the approximation error remains ultimately bounded.  
 \end{remark}}

\begin{algorithm}[t]
\caption{Model-Free Optimal Tracking Control}
\label{alg:Data_OT}
\begingroup
\setlength{\abovedisplayskip}{2pt}
\setlength{\belowdisplayskip}{2pt}
\setlength{\abovedisplayshortskip}{2pt}
\setlength{\belowdisplayshortskip}{2pt}
\begin{algorithmic}[1]
\setlength{\itemsep}{1pt}
\setlength{\parsep}{0pt}
\setlength{\parskip}{0pt}

\STATE {\bf Initialization:} ${K}_0=0$, $\alpha_0>0$, $\gamma>\max\{\alpha_0,\sigma(\mathcal{L}_{[0;\mathcal{S}]})+\alpha_0\}$, $\eta\in(0,1)$, $\varepsilon>0$, $i=0$. 
\STATE {\bf Data collection:} Inject $\ell(t)$ into the input channel, {\sl i.e.}, $u(t)=\ell(t)$, and construct the data-driven matrices until the rank conditions \eqref{eq:rank1} and \eqref{eq:rank2} are satisfied. 
\REPEAT
\STATE $i \leftarrow i+1$
\STATE Solve ${\theta}_i$ from
\begin{equation}\label{eq:int_P_i_DB}
\begin{aligned}
{\theta}_i = -{\Psi}_{i-1}^{\dagger} {\mathcal{I}}_{ \chi}^{\top} \mathrm{vech}(\|{K}_{i-1}\|_R^2 + \Theta).
\end{aligned}
\end{equation}
\STATE Update ${K}_i$ by
\begin{equation}\label{eq:int_K_i_DB}
{K}_i=(R+{\Lambda}_i)^{-1}{M}_i.
\end{equation}
\STATE Update ${\alpha}_i$ by
\begin{equation}\label{eq:int_alpha_i_DB}
{\alpha}_i=\mathcal{R}({\alpha}_{i-1},{P}_i,{K}_i).
\end{equation}
\UNTIL{$\alpha_{i}\geq\gamma$}
\REPEAT
\STATE $i \leftarrow i+1$
\STATE Solve ${\theta}_i$ from
\begin{equation}\label{eq:P_i_DB}
{\theta}_i=- {\Phi}_{i-1}^{\dagger}({\mathcal{I}}_{\chi}^{\top}\mathrm{vech}(\|{K}_{i-1}\|_R^2+\mathcal{I}_{\zeta}^{\top}\mathrm{vech}(Q)).
\end{equation}
\STATE Update ${K}_{i}$ by \eqref{eq:int_K_i_DB}.
\UNTIL{$\|{P}_i-{P}_{i-1}\|_F\leq\varepsilon$}
\STATE $P^*= {P}_i$, $K^*= {K}_i$.
\STATE Solve ${\vartheta}^*$ from
\begin{equation}\label{eq:FF_DB}
{\vartheta}^* = {\Xi}^{\dagger} \mathcal{I}_{y_d{\bar{\zeta}}}^{\top}\mathrm{vec}\left(Q\right). 
\end{equation}
\end{algorithmic}
\endgroup
\end{algorithm}

\section{Numerical Example}\label{sec4}
\begin{table*}[htbp]
\centering
\caption{Output trajectories of the desired system with different $H_d$ and their corresponding estimated FF gains.}
\label{tab:Hd}
\setlength{\tabcolsep}{4pt}   
\footnotesize                  
\begin{tabular}{@{} c | c | c | c|c | c | c | c |c@{}}
\toprule
Case 
  & 1
  & 2 
  & 3 
  & 4
  & 5 
  & 6 
  & 7
  & 8  
  \\
\midrule
$H_d$ 
&$[1,0,0]$
&$[2,0,0]$
&$[3,0,0]$
&$[3,1,-1]$
&$[1,1,0]$
&$[1,0,1]$
&$[0,1,1]$
&$[0,1,0]$ 
\\
\midrule
$y_d(t)$ 
 & \begin{minipage}{0.1\textwidth}\centering\includegraphics[width=\linewidth]{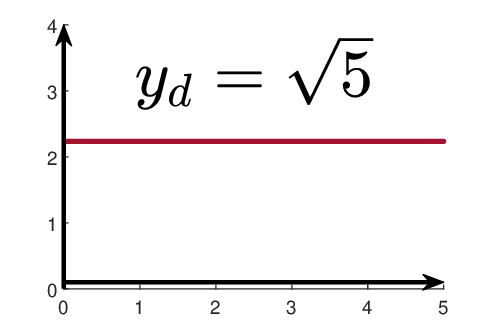}\end{minipage}
  & \begin{minipage}{0.1\textwidth}\centering\includegraphics[width=\linewidth]{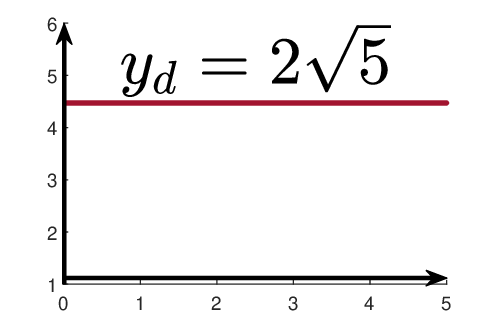}\end{minipage}
& \begin{minipage}{0.1\textwidth}\centering\includegraphics[width=\linewidth]{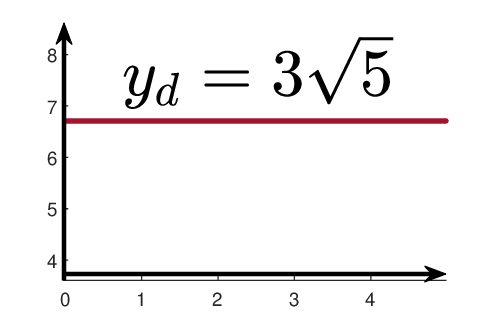}\end{minipage}
  & \begin{minipage}{0.1\textwidth}\centering\includegraphics[width=\linewidth]{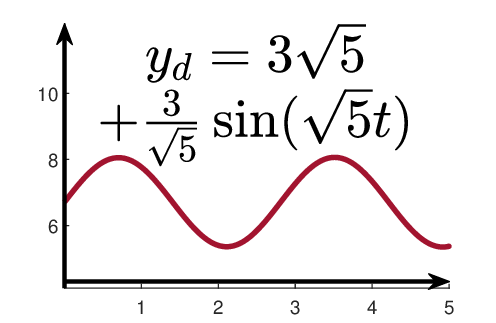}\end{minipage}
 & \begin{minipage}{0.1\textwidth}\centering\includegraphics[width=\linewidth]{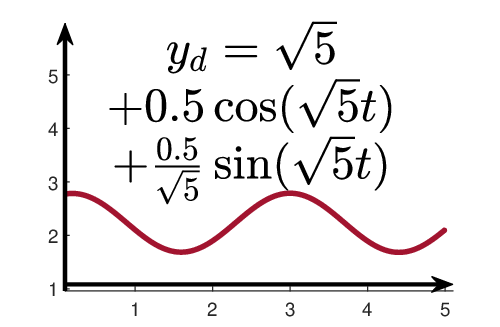}\end{minipage}
& \begin{minipage}{0.1\textwidth}\centering\includegraphics[width=\linewidth]{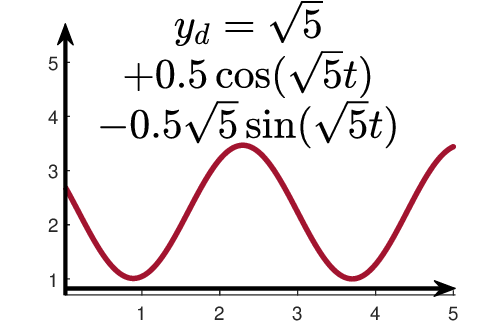}\end{minipage}
& \begin{minipage}{0.1\textwidth}\centering\includegraphics[width=\linewidth]{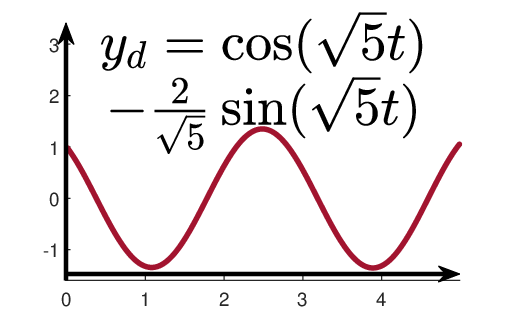}\end{minipage}
& \begin{minipage}{0.1\textwidth}\centering\includegraphics[width=\linewidth]{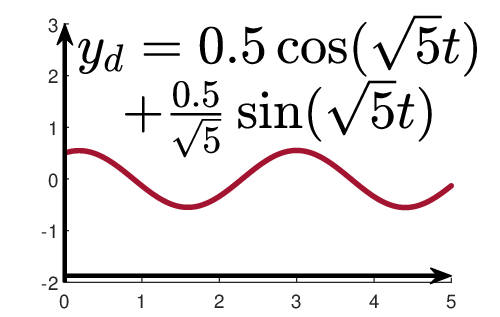}\end{minipage}\\
\midrule
$\hat{F}^*$ 
&$\left[\begin{smallmatrix}-29.898\\0.189\\0.07\end{smallmatrix}\right]^{\top}$
&$\left[\begin{smallmatrix}-59.796\\0.379\\0.14\end{smallmatrix}\right]^{\top}$
&$\left[\begin{smallmatrix}-89.695\\0.568\\0.21\end{smallmatrix}\right]^{\top}$
&$\left[\begin{smallmatrix}-89.676\\-60.932\\16.745\end{smallmatrix}\right]^{\top}$
&$\left[\begin{smallmatrix}-29.887\\-23.908\\-7.42\end{smallmatrix}\right]^{\top}$
&$\left[\begin{smallmatrix}-29.905\\37.592\\-23.954\end{smallmatrix}\right]^{\top}$
&$\left[\begin{smallmatrix}0.004\\13.305\\-31.514\end{smallmatrix}\right]^{\top}$
&$\left[\begin{smallmatrix}0.011\\-24.098\\-7.49\end{smallmatrix}\right]^{\top}$\\
\midrule
\end{tabular}
\end{table*}
Consider a stochastic spring–mass–damper system subject to multiplicative noise, with system matrices 
\[
A=\begin{bmatrix}0&1\\-5&-0.5\end{bmatrix},
B=\begin{bmatrix}0\\1\end{bmatrix},
C=\begin{bmatrix}0.1&0.2\\0.2&0.3\end{bmatrix},
D=\begin{bmatrix}0\\0.1\end{bmatrix},
\]
\(H=[1\;0]\), and initial condition
\(x_0=[0\;0]^\top\). The parameters of the exosystem are given by
\[
A_d=\begin{bmatrix}
0&0&0\\
0&0&1\\
0&-5&0
\end{bmatrix},\qquad
x_{d0}=\begin{bmatrix}\sqrt{5}\\0.5\\0.5\end{bmatrix}.
\]
Different reference trajectories can be produced by selecting different output
matrices \(H_d\). In this example, eight reference system outputs are considered, as
listed in Table I. The weighting matrices are chosen as $Q=10$ and $R=0.01$.

To implement Algorithm \ref{alg:Data_OT}, the parameters are chosen as $\gamma=1$, $\alpha_0=0.1$, $\eta=0.95$,  $\Theta = \mathrm{diag}([10,10])$, and $\varepsilon=10^{-3}$. 
The initial FBG is $K_0=0$. During data collection, the discounted input is selected as $\nu(t)=10\sum_{j=1}^{50}\sin(\omega_j t)$, where $\omega_j$ are randomly selected from $[-100,100]$. The data are collected over $[t_1,t_l+T]=[0,5.1]$ s, with sampling period $T_s=0.001$ s and integration window length $T=0.1$ s.

\begin{figure}[!hbt]
\centering
\includegraphics[scale=0.4]{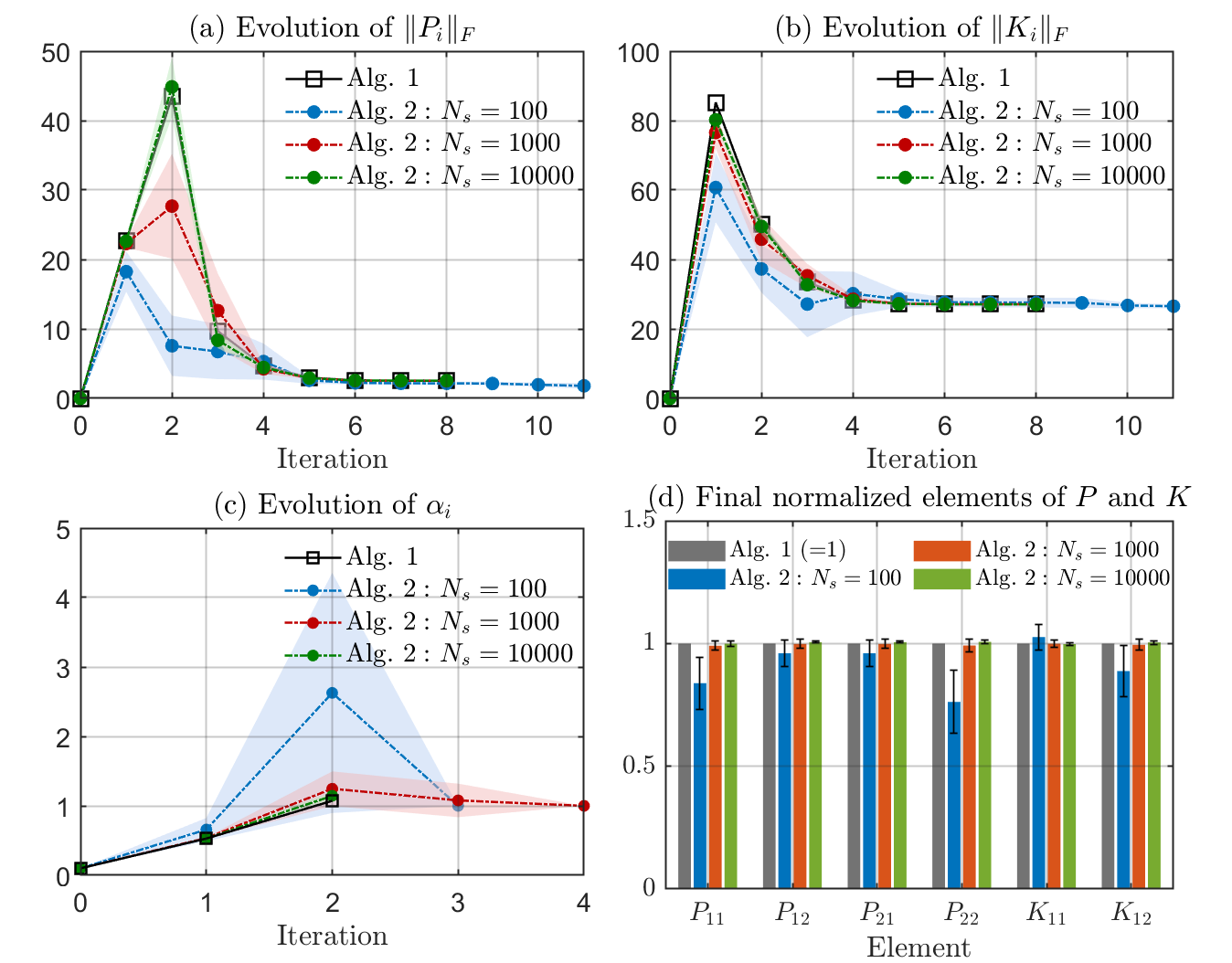}
\caption{Convergence of the learned parameters.}\label{rf1}
\end{figure}

We first evaluate the learning performance of the off-policy RL component in Algorithm \ref{alg:Data_OT}.
\black{Algorithm \ref{alg:Model_CPI} is applied to obtain the optimal solution $(P^*,K^*)$. The resulting sequences are used as iteration-wise benchmarks in Fig. \ref{rf1} (a)-(c), and $(P^*,K^*)$ is used to assess the accuracy of the final estimates in Fig. \ref{rf1}(d).}
\black{To examine the influence of finite-sample approximation, Algorithm \ref{alg:Data_OT} is independently executed over $20$ Monte Carlo trials for each $N_s\in\{100,1000,10000\}$. Figs. \ref{rf1} (a)-(c) show the evolution of $\|P_i\|_F$, $\|K_i\|_F$, and $\alpha_i$, respectively. The solid curves represent the empirical means over $ 20$ trials, and the corresponding error bars indicate one standard deviation. For all three sample sizes, the learned parameters converge to neighborhoods of the model-based iteration sequences. Moreover, increasing $N_s$ reduces both the estimation bias and the variability across trials. 
Fig. \ref{rf1} (d) compares the normalized componentwise estimates of $P^*$ and $K^*$ obtained at convergence. The same trend is observed, that is larger values of $N_s$ yield estimates that are more tightly concentrated around the model-based optimal solution. These results confirm that Algorithm \ref{alg:Data_OT} provides a consistent finite-sample implementation of the expectation-based learning equations, with improved accuracy as more trajectories are used.}
\black{The average computation times per trial for $N_s=100,1000,$ and $10000$ are $0.255$ s, $1.153$ s, and $8.701$ s, respectively, on a laptop with $32$ GB RAM and $2.6$ GHz Intel Core i9 processor.}
\begin{figure}[!htb]
\centering
\includegraphics[scale=0.30]{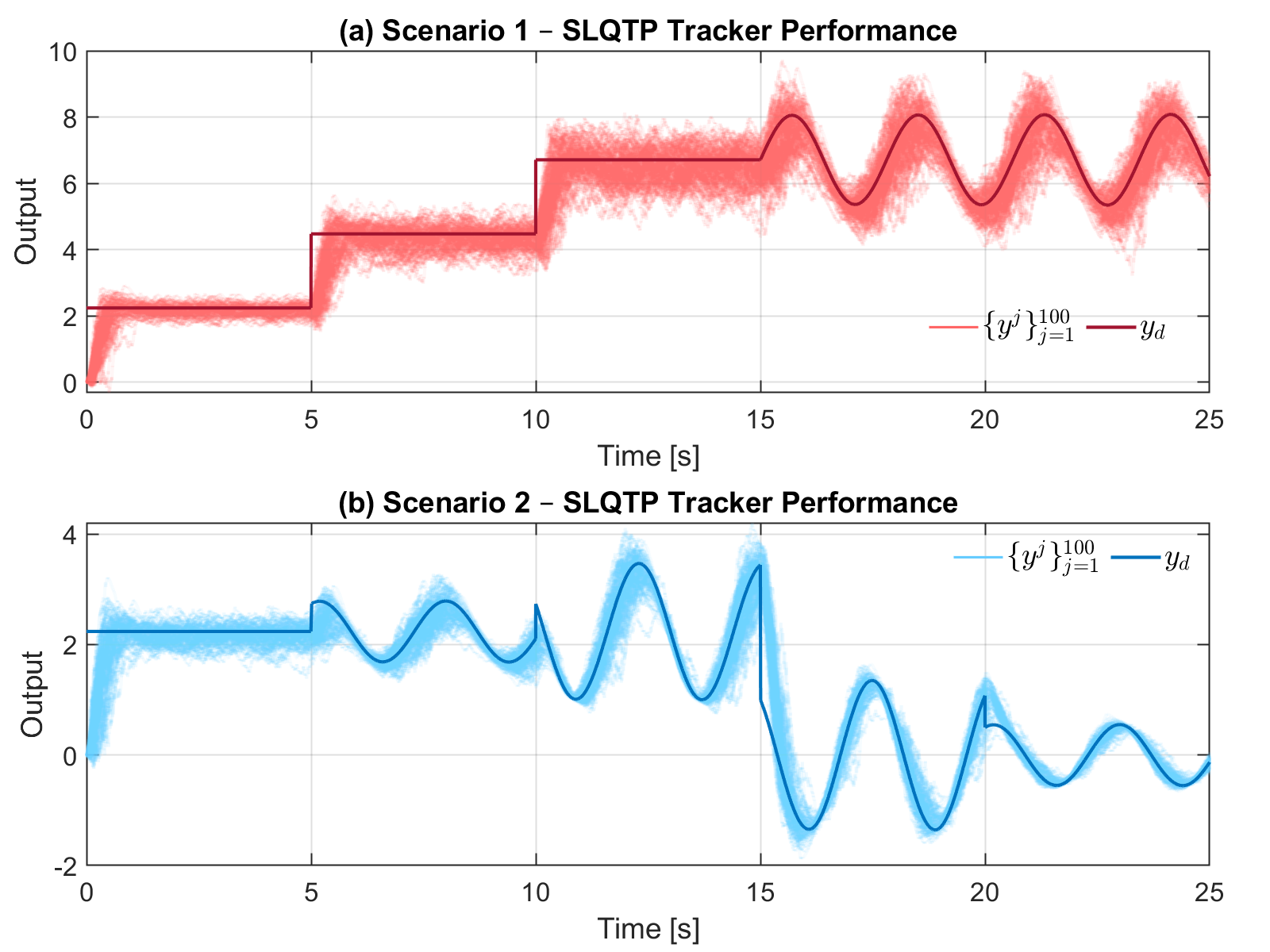}
\caption{Performance of the learned optimal tracker.}\label{rf2}
\end{figure}

We next evaluate the tracking performance of the learned tracker. One FBG learned using $N_s=1000$ is selected, and the corresponding FFGs for the
eight reference signals are computed from \eqref{eq:FF_DB}.
The results are summarized in Table \ref{tab:Hd}.
Two switching reference scenarios are considered.
In Scenario 1, the reference switches as $y_d^1\rightarrow y_d^2\rightarrow y_d^3\rightarrow y_d^4$, where the first three segments last $5$ s and the final segment lasts $10$ s. In Scenario 2, the reference switches as $y_d^1\rightarrow y_d^5\rightarrow y_d^6\rightarrow y_d^7\rightarrow y_d^8$, with each segment lasting $5$ s.
For each scenario, $100$ independent closed-loop output trajectories are generated from the same initial condition, with \(y^j\) denoting the \(j\)-th sampled trajectory in Fig. \ref{rf2}. 
Following each reference switch, the output trajectories remain concentrated around the desired signal, and the tracker rapidly adapts to changes. 
These results demonstrate that reference trajectories generated by the same underlying dynamics can be effectively tracked using the corresponding optimal controllers, which share a common FBG while employing trajectory-specific FFGs.
\begin{figure}[!htb]
\centering
\includegraphics[scale=0.30]{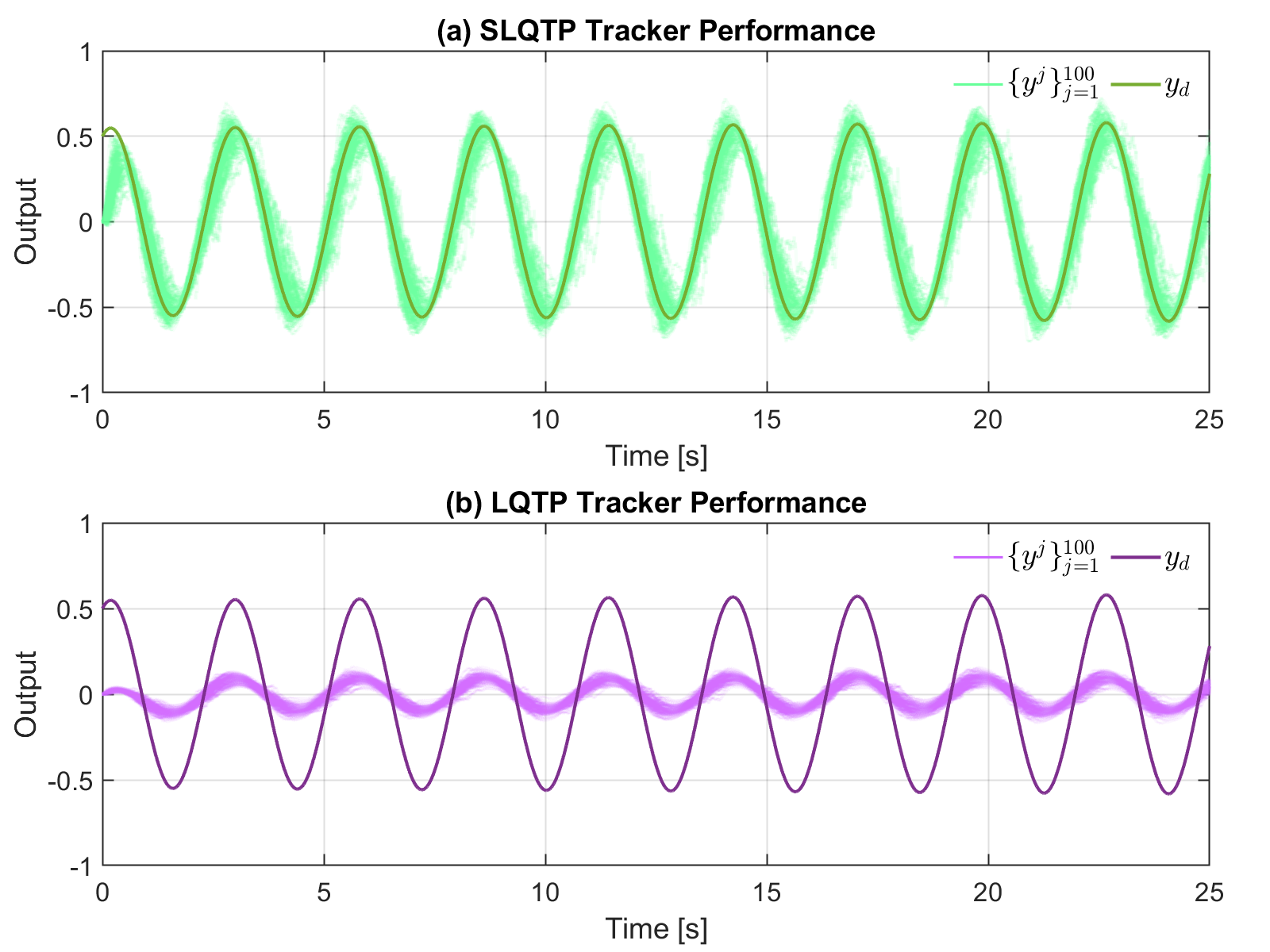}
\caption{Performance of the learned optimal tracker compared with the one of LQTP.}\label{rf3}
\end{figure}

Finally, the proposed SLQTP controller is compared with a  deterministic LQTP controller designed by neglecting the diffusion dynamics, {\sl i.e.,} $C=0$ and $D=0$. Both controllers are applied to the original stochastic system, and reference case 8 is used for comparison. As shown in Fig. \ref{rf3}, the SLQTP tracker significantly outperforms the LQTP tracker, maintaining output trajectories close to the desired reference. This demonstrates the importance of incorporating multiplicative noise into stochastic tracking control design.

\section{Conclusion}\label{sec5}
This paper developed a bootstrap policy iteration-based optimal tracking framework for continuous-time stochastic linear systems with multiplicative noise.
The proposed method characterizes the optimal tracker by separating the computation of the optimal feedback and feedforward gains.
In addition, the B-PI algorithm removes the restrictive requirement for an initial stabilizing policy.
The resulting off-policy RL algorithm can also be applied to the stochastic linear quadratic regulation problem.

\appendix
\subsection{Proof of Lemma \ref{lem:LE-discount}} \label{sec:proof_of_lemma_LE-discount}
{\noindent 1)}. Given $K\in\mathcal{Z}_{(\alpha,\gamma)}$, we first consider \eqref{Lya-alpha} with $\mathcal{Q}=\Theta$. According to \cite[Theorem 3.2.3]{sun2020stochastic}, this equation admits a unique solution $P$ given by
\begin{equation}\label{eq:Pexpress}
P=\mathbb{E}\left[\int_0^{\infty}X(\tau)^{\top}\left(\|K\|_R^2+\Theta\right)X(\tau)\mathrm{d}\tau\right],
\end{equation}
where $\{X(t);t\geq0\}$ satisfies the matrix SDE $\mathrm{d}X = (A_{(\alpha,\gamma)} -BK)X\mathrm{d}t+(C-DK)X\mathrm{d}w$, $X(0)=I_n$. 

Because $K\in\mathcal{Z}_{(\alpha,\gamma)}$ also implies exponential mean square stability of the above system due to time invariance (see \cite{khasminskii2012stochastic}), and since $R,\Theta>0$, it follows from \eqref{eq:Pexpress} that $0<{P}<cI$ for some finite constant $c>0$.

To establish that $\alpha'>\alpha$ and $K'\in\mathcal{Z}_{(\alpha',\gamma)}$, we substitute the definition of $K'$ into (\ref{Lya-alpha}) and complete the squares to obtain
\begin{equation}\label{eq:0717-2}
\begin{aligned}
\mathcal{L}_{[K';\mathcal{S}_{(\alpha',\gamma)}]}({P})+\|K'-{K}\|_{\Upsilon}^2+\Theta'=(\alpha'-\alpha){P},
\end{aligned}
\end{equation}
where $\Theta'=\|K'\|_R^2+\Theta>0$ and $\Upsilon=R+D^{\top}PD>0$. 
By using $\alpha'=\mathcal{R}(\alpha,P,K',\Theta)$, we have $\alpha'-\alpha = \eta {\lambda_{\min}(\Theta')}/{\lambda_{\max}(P)}$. Since $\eta\in(0,1)$, $\Theta'>0$, and $P>0$, it clearly implies that $\alpha'-\alpha>0$. Moreover, we get  
\begin{equation}\label{eq:0717-1}
\Theta'\geq \lambda_{\min}(\Theta')I_n>(\alpha'-\alpha)\lambda_{\max}(P)I_n\geq (\alpha'-\alpha)P.
\end{equation} 
Combing \eqref{eq:0717-2} and \eqref{eq:0717-1} gives 
$\mathcal{L}_{[K';\mathcal{S}_{(\alpha',\gamma)}]}({P})<0$.
By \cite[Theorem 1]{rami2000linear}, we conclude that $\sigma(\mathcal{L}_{[K';\mathcal{S}_{(\alpha',\gamma)}]})\subset\mathbb{C}^-$, and thus $K'\in\mathcal{Z}_{(\alpha',\gamma)}$. \\
{\noindent 2)}. Since $\mathcal{S}_{(\gamma,\gamma)}=\mathcal{S}$, we subtract (\ref{eq:P*}) from \eqref{Lya-alpha} with $\mathcal{Q}=\|H\|_Q^2$ and $\alpha=\gamma$ to obtain
\begin{equation}\label{eq:0717-3}
\begin{aligned}
\mathcal{L}_{[{K};\mathcal{S}]}({P}-{P}^*)+\|{K}-{K}^*\|_{\Upsilon^*}^2=0,
\end{aligned}
\end{equation}
where $\Upsilon^*=R+D^{\top}P^*D>0$. Together with $K\in\mathcal{Z}_{(\gamma,\gamma)}$, this equation implies that ${P}\geq {P}^*$. 

Next, we aim to show that  $K'\in\mathcal{Z}_{(\gamma,\gamma)}$. Rewriting \eqref{eq:0717-3} using $K'=\mathcal{K}(P)$ gives
\begin{equation}
\begin{aligned}
\mathcal{L}_{[K';\mathcal{S}]}({P}-{P}^*)+\Delta K=0,
\end{aligned}
\end{equation}
where $\Delta K := \|K'-{K}^*\|_{\Upsilon^*}^2+\|K'-{K}\|_{\Upsilon}$. According to \cite[Theorem 3.2]{zhang2008generalized}, since $P-P^*\geq0$ and $\Delta K\geq0$ hold, it is sufficient to prove the exact detectability of the pair $[A-BK',\Delta K|C-DK']$ to conclude that $K'\in\mathcal{Z}_{(\gamma,\gamma)}$. 

Suppose, to the contrary, that the pair is not exactly detectable.
Then, by the stochastic Popov–Belevitch-Hautus criterion (\cite[Theorem 3.1]{zhang2008generalized}), there exists a nonzero symmetric matrix $X\neq0$ such that 
$\mathcal{L}_{[{K}';\mathcal{S}]}(X)=\lambda X,~Re(\lambda)\geq0$ and $\Delta KX=0$.
This implies that $K'X={K}X$ and hence $\mathcal{L}_{[K';\mathcal{S}]}(X)=\mathcal{L}_{[{K};\mathcal{S}]}(X)=\lambda X$.
However, since $K\in\mathcal{Z}_{(\gamma,\gamma)}$, it holds that $\sigma(\mathcal{L}_{[K;\mathcal{S}]})\subset\mathbb{C}^-$, which contradicts $\mathrm{Re}(\lambda)\geq0$.
Thus, the exact detectability of $[A-BK',\Delta K|C-DK']$ holds.
This completes the proof. \hfill $\square$

\subsection{Proof of Lemma \ref{lem:rank}}\label{lempf:rank}
We prove the result by contradiction. Suppose there exists a nonzero vector 
$\mathbf{y} =[\mathbf{y}_1^{\top},\mathbf{y}_2^{\top},\mathbf{y}_3^{\top}]^{\top}$, such that $\varPhi(\alpha,K)\mathbf{y}={0}$. Let $\mathbf{y}_1=\mathrm{vech}(Y_1)\in\mathbb{R}^{\frac{n}{2}(n+1)}$, $\mathbf{y}_2=\mathrm{vec}(Y_2)\in\mathbb{R}^{mn}$ with $Y_2\in\mathbb{R}^{m\times n}$, and $\mathbf{y}_3=\mathrm{vech}(Y_3)\in\mathbb{R}^{\frac{m}{2}(m+1)}$. Then by using the definition of $\varPhi(\alpha,K)$   and \eqref{eq:psi}, we have
 \begin{equation}\label{eq:pflem2eq1}
 \begin{aligned}
0&=\Delta_{\chi}^{\top}\mathbf{y}_1+\mathcal{I}_{\chi}^{\top}\mathrm{vech}\big((\alpha-\alpha_0)Y_1-K^{\top}Y_2\\
&~~~~-Y_2^{\top}K+K^{\top}Y_3K\big)-2\mathcal{I}_{\chi \nu}^{\top}\mathbf{y}_2-\mathcal{I}_{\nu}^{\top}\mathbf{y}_3.
\end{aligned}
\end{equation}
Next, applying  It\^{o}'s formula to $\chi^{\top}Y_1\chi$ along with \eqref{sys:xtilde} yields
\begin{equation*}
\begin{aligned}
\mathrm{d}\left(\chi^{\top}Y_1\chi\right) & = \chi^{\top} F_1(Y_1)\chi\mathrm{d}t+2\nu^{\top}F_2(Y_1)\chi\mathrm{d}t\\
&~~+\nu^{\top}F_3(Y_1)\nu\mathrm{d}t+2\chi^{\top}Y_1\left(C\chi+D\nu\right)\mathrm{d}w,
\end{aligned}
\end{equation*}
where $ F_1(Y_1) =  \mathcal{L}_{[K;\mathcal{S}_{(\alpha,\gamma)}]}(Y_1)-(\alpha-\alpha_0) Y_1+K^{\top}(B^{\top}Y_1+D^{\top} Y_1 C)+ (Y_1 B + C^{\top} Y_1 D) K-K^{ \top} D^{ \top} Y_1 DK$, $F_2(Y_1) = B^{\top}Y_1+D^{\top}Y_1C$, and $F_3(Y_1) = D^{\top}Y_1D.$ Integrating the preceding identity over $[t_k,t_k+T]$, $k=1,\cdots,l$, taking expectations, and expressing the resulting relations in matrix form using the Kronecker product representation, we obtain
\begin{equation}\label{eq:pflem2eq2}
\Delta_{\chi}^{\top}\mathbf{y}_1 = \mathcal{I}_{\chi}^{\top}\mathrm{vech}\left(F_1\right) + 2\mathcal{I}_{\chi \nu}^{\top}\mathrm{vec}\left(F_2\right) + \mathcal{I}_{\nu}^{\top}\mathrm{vech}\left(F_3\right).
\end{equation} 
Substituting \eqref{eq:pflem2eq2} into \eqref{eq:pflem2eq1} and collecting like terms yield
\begin{equation}\label{43}
\mathcal{I}^{\top}[
\mathrm{vech}(N_1)^{\top},
\mathrm{vec}(N_2)^{\top},
\mathrm{vech}(N_3)^{\top}]^{\top}={0},
\end{equation}
where $N_1=\mathcal{L}_{[K;\mathcal{S}_{(\alpha,\gamma)}]}(Y_1)+K^{\top}N_2+N_2^{\top}K-K^{\top}N_3K$, $N_2=B^{\top}Y_1+D^{\top}Y_1C-Y_2$, $N_3=D^{\top}Y_1D-Y_3$.

By Assumption \ref{assume:rank1}, we have $\mathcal{I}\mathcal{I}^{\top}>0$. It follows from (\ref{43}) that $N_1={0}$, $N_2={0}$, and $N_3={0}$. In particular, $N_1=0$ yields $\mathcal{L}_{[K;\mathcal{S}_{(\alpha,\gamma)}]}(Y_1)={0}$, together with $K\in\mathcal{Z}_{(\alpha,\gamma)}$, which gives that $Y_1={0}$. Substituting this into  $N_2=0$ and $N_3=0$ further gives $Y_2={0}$ and $Y_3={0}$. Consequently, we have $\mathbf{y}={0}$, which contradicts the assumption that  $\mathbf{y}\neq0$. Hence, the matrix $\varPhi(\alpha,K)$ has full column rank. \hfill $\square$

\subsection{Proof of Theorem \ref{thm:DPI}}\label{sec:proof_of_theorem_ref_thm_dpi}
{1)}. We first prove by mathematical induction that $0<P_i< cI$, $K_{i}\in\mathcal{Z}_{(\alpha_{i},\gamma)}$, and $\alpha_i>\alpha_{i-1}$ hold for all $1\leq i\leq \mathbb{I}$, where $c$ is a finite positive constant.\\
{\romannumeral1}). Suppose $i=1$. Since $\gamma>\sigma(\mathcal{L}_{[0;\mathcal{S}]})+\alpha_0$, it follows from Lemma \ref{lem:alpha_init} that $K_{0}=0\in\mathcal{Z}_{(\alpha_0,\gamma)}$. Then, by applying Lemma \ref{lem:LE-discount} (Statement 1), we conclude that \eqref{eq:int_P_i} admits a unique solution $0<P_1<c I$ such that   $K_1\in\mathcal{Z}_{(\alpha_1,\gamma)}$ and $\alpha_1>\alpha_0$.\\
{\romannumeral2}). Assume for some $1<i<\mathbb{I}$, $K_{i}\in\mathcal{Z}_{(\alpha_{i},\gamma)}$ holds. Apply Lemma \ref{lem:LE-discount}(Statement 1) again at step $i+1$, we have that  \eqref{eq:int_P_i} admits a unique solution $0<P_{i+1}<c I$, and the updated FBG $K_{i+1}$ belongs to $\mathcal{Z}_{(\alpha_{i+1},\gamma)}$ with $\alpha_{i+1}>\alpha_{i}$. 

Therefore, the sequence $\{\alpha_i\}$ is strictly increasing and bounded above by $\gamma$. By construction, the stopping index $\mathbb{I}$ satisfies that $\gamma\leq \alpha_{\mathbb{I}}$ and $\alpha_{i}<\gamma$ holds for all $i<\mathbb{I}$. Since each $P_i$ is bounded, there exist positive constants $0<c_0,c_1<\infty$ such that  $\max_{1\leq i\leq \mathbb{I}}\lambda_{\max}(P_i)\leq c_1$ and $\min_{1\leq i\leq \mathbb{I}}\lambda_{\min}(\|K_i\|_R^2+\Theta)\geq c_0$. Applying these bounds to \eqref{eq:int_alpha_i}, we obtain $\alpha_i-\alpha_{i-1}\geq \eta c_0/c_1$. A summation from $i=1$ to $i=\mathbb{I}-1$ gives $\gamma-\alpha_0>\alpha_{\mathbb{I}-1}-\alpha_0\geq (\mathbb{I}-1)\frac{\eta c_0}{c_1},$ 
which implies that
$\mathbb{I}<1 + \frac{(\gamma-\alpha_0)c_1}{\eta c_0}<\infty$.
This proves the finiteness of $\mathbb{I}$.
Then, at iteration $i=\mathbb{I}$, \eqref{eq:int_P_i} becomes
\begin{equation*}
\begin{aligned}
\mathcal{L}_{[K_{\mathbb{I}-1};\mathcal{S}_{(\alpha_{\mathbb{I}-1},\gamma)}]}(P_{\mathbb{I}}) +\|K_{\mathbb{I}-1}\|_R^2+\Theta=0.
\end{aligned}
\end{equation*}
Express it in terms of the original system $\mathcal{S}$ and ${K}_{\mathbb{I}}$ as
\begin{equation*}
\begin{aligned}
\mathcal{L}_{[K_{\mathbb{I}};\mathcal{S}]}(P_{\mathbb{I}}) = -\|K_{\mathbb{I}}-K_{\mathbb{I}-1}\|_{\Upsilon_{\mathbb{I}}}^2-\Theta_{\mathbb{I}}-(\alpha_{\mathbb{I}}-\gamma)P_{\mathbb{I}},
\end{aligned}
\end{equation*}
where $\Theta_{\mathbb{I}}:=\|K_{\mathbb{I}}\|_R^2+\Theta+(\alpha_{\mathbb{I}-1}-\alpha_{\mathbb{I}})P_{\mathbb{I}}$. The positive definiteness of $\Theta_{\mathbb{I}}$ directly follows from \eqref{eq:int_alpha_i} and the facts that $P_{\mathbb{I}}>0$ and $\alpha_{\mathbb{I}}>\alpha_{\mathbb{I}-1}$.
Finally, since $\alpha_{\mathbb{I}}\geq\gamma$, by \cite[Theorem 1]{rami2000linear}, it can be immediately concluded that $K_{\mathbb{I}}\in\mathcal{Z}_{(\gamma,\gamma)}$.

{2)} Next, we verify that for all $i\geq\mathbb{I}+1$,  $P_i\geq P^*$ and $K_{i}\in\mathcal{Z}_{(\gamma,\gamma)}$ by mathematical induction.\\
   {\romannumeral1}). Suppose $i=\mathbb{I}+1$. 
    Since $K_{\mathbb{I}}\in\mathcal{Z}_{(\gamma,\gamma)}$, based on Lemma \ref{lem:LE-discount}(Statement 2), we know that \eqref{eq:P_i} admits a unique solution $P_{\mathbb{I}+1}\geq P^*$ such that $K_{\mathbb{I}+1} \in \mathcal{Z}_{(\gamma,\gamma)}$. \\
    {\romannumeral2}).Assume for some $i>\mathbb{I}+1$, $K_{i}\in\mathcal{Z}_{(\gamma,\gamma)}$ holds. Apply Lemma \ref{lem:LE-discount}(Statement 2) again at step $i+1$, the solutions to \eqref{eq:P_i} and \eqref{eq:int_K_i} satisfy  $P_{i+1}>0$ and $K_{i+1}\in\mathcal{Z}_{(\gamma,\gamma)}$. 

Finally, \eqref{eq:P_i} and \eqref{eq:int_K_i} yield the recursive identity
\begin{equation*} 
\begin{aligned}
\mathcal{L}_{[K_{i};\mathcal{S}]}(P_{i}-P_{i+1})+\|K_{i}-K_{i-1}\|_{{R}}^{2}=0,~\forall i\geq \mathbb{I}+1.
\end{aligned}
\end{equation*}
Since $\sigma(\mathcal{L}_{[K_{i};\mathcal{S}]})\subset \mathbb{C}^-$, it follows that $P_{i}\geq P_{i+1}$ for all $i\geq\mathbb{I}+1$. Combining this monotonicity property with the previously established lower bound $P_i\geq P^*$, we conclude that the sequence $\{P_i\}_{i=\mathbb{I}+1}^{\infty}$ has a limit $P_{\infty}$ such that $\lim_{i\rightarrow\infty}P_i=P_{\infty}$, and then  $\lim_{i\rightarrow\infty}K_i=\mathcal{K}(P_{\infty}):=K_{\infty}$. Passing to the limit in \eqref{eq:P_i} and \eqref{eq:int_K_i}, we find that $(P_{\infty},K_{\infty})$ satisfies the same SARE as the stabilizing pair $(P^*,K^*)$. Therefore, under Assumption \ref{A1}, we have $P_{\infty}=P^*$ and $K_{\infty}=K^*$. \hfill $\square$

 \bibliographystyle{IEEEtran}
\bibliography{Refxzh_2023.bib}

\end{document}